\documentclass[10pt,a4paper]{article}

\usepackage[T2A, T1]{fontenc} 
\usepackage[english, russian]{babel}

\usepackage[affil -it]{authblk}

\usepackage{graphicx}
\usepackage{amsmath,amsfonts,amssymb,textcomp}
\usepackage{color}
\usepackage{xcolor}

\usepackage{amsmath}

\usepackage[sorting=none, style=phys, backend=biber, uniquename=init, giveninits=true,  maxbibnames=5, minbibnames=5]{biblatex}

\bibliography{references}

\usepackage[colorlinks]{hyperref}
\usepackage[colorinlistoftodos, textsize=tiny, textwidth=110,]{todonotes}
\setuptodonotes{fancyline}

\usepackage{array,multirow,tabularx, makecell}
\usepackage{adjustbox}

\title{\textbf{Interaction of accelerator neutrinos with energies up to 55 MeV with~${}^{127}$I nuclei}}
\author[1*]{Yu. S. Lutostansky}
\author[1,2**]{A. N. Fazliakhmetov}
\author[1]{V. N. Tikhonov}
\author[2]{G. A. Koroteev}
\author[1]{N. A. Belogortseva}
\author[1]{N. V. Klochkova}
\author[1]{A. Yu. Lutostansky}
\author[1]{A. P. Osipenko}
\author[1]{E. Yu. Zemskov}

\affil[1]{ National Research Centre ''Kurchatov Institute'', Moscow, Russia}
\affil[2]{Institute for Nuclear Research of Russian Academy of Sciences, Moscow, Russia}

\affil[ ]{ }
\affil[ ]{E-mail:}
\affil[*]{ lutostansky@yandex.ru}
\affil[**]{ fazliakhmetov@phystech.edu}

\begin{document}

\selectlanguage{english}

\maketitle

\begin{abstract}
The interaction of neutrinos with an energy of up to 55~MeV from the Spallation Neutron Source (SNS) accelerator with a perspective ${}^{127}$I detector at the Oak Ridge National Laboratory (United States) has been studied. 
The resonance structure of the charge-exchange strength function $S(E)$ has been calculated taking into account high-lying resonances, and the effect of this structure on the cross section $\sigma(E)$ for the accelerator neutrino capture by the ${}^{127}$I nucleus has been examined. 
The influence of the Gamow–Teller resonance GTR-1 and the second new higher resonance GTR-2 on the energy dependence of the cross section $\sigma(E)$ has been analyzed. 
The effect of the high-lying analog resonance AR-2 has also been taken into account for the first time. 
It has been found that the contributions of GTR-1 to the calculated cross section $\sigma(E)$ are from 60\% to $\approx$80\%, and GTR-2 about 12\%, and AR-2 $\le$ 10\%, respectively. 
The contribution of high-lying resonances to the $(\nu, n)$ and $(\nu, 2n)$ neutrino capture cross sections with neutron emission and the formation of the ${}^{126}$I and ${}^{125}$I isotopes, respectively, has been analyzed. 
The comparison of the cross sections for the interaction of accelerator neutrinos calculated by different methods with experimental data has shown coincidence at energies below the neutron separation threshold $E_x < S_{1n}$ and strong discrepancy at higher energies, which is difficult to explain. 
A new measurement of cross sections at energies $E_x > S_{1n}$ is needed.
\end{abstract}

\section{Introduction}
For the neutrino registration, detectors containing various isotopes are widely used, the interaction cross-sections of which with neutrinos $\sigma(E)$ are usually well known from experimental data and calculations. 
This is typically associated with neutrinos with the energies up to 20--30~MeV, before neutron emission from the target daughter nuclei has begun. 
For higher energies, experimental data on the cross section $\sigma(E)$ for detector nuclei are usually absent and calculations of high-lying charge-exchange excitations for the strength function $S(E)$ of the corresponding detector isotope have not been performed.

In this paper, we study the interaction of neutrinos with energies up to 55~MeV with the promising detector based on the isotope ${}^{127}$I. 
Neutrino detection is based on the iodine-xenon radiochemical method, based on the reaction
\begin{equation}
	\label{eq:1}
	\nu_{e} + {}^{127}\mathrm{I} \to e^{-} + {}^{127}\mathrm{Xe}
\end{equation}
which is similar to chlorine-argon, but has a lower detection threshold, which increases the neutrino capture cross-section $\sigma(E)$.

A new ${}^{127}$I neutrino detector has been recently used by the COHERENT Collaboration to measure the neutrino capture cross section on a neutrino flux from the Spallation Neutron Source (SNS) accelerator. 
The capture cross section of electron neutrinos with an energy of up to 55~MeV from the decay of $\pi^{+}$ mesons from the SNS accelerator was measured for the first time at the Oak Ridge National Laboratory \cite{PhysRevLett.131.221801}. In the early 2000s, the neutrino capture cross section was measured out on a prototype of the iodine radiochemical solar neutrino detector at the Los Alamos National Laboratory \cite{PhysRevC.68.054613}. 
In this experiment, only transitions to bound states of ${}^{127}$Xe were considered, and the processes associated with the excitation of the ${}^{127}$Xe nucleus above the separation energy of one or two neutrons were not investigated.

The use of ${}^{127}$I as a radiochemical detector for solar and supernova neutrinos was first proposed by W. C. Haxton in 1988~\cite{PhysRevLett.60.768}. In a recent paper the authors suggested that such a detector could be used to distinguish between different solar models \cite{LutostanskyEtAl:2020_JETP_lett, LUTOSTANSKY2022136905}. 
The calculations for accelerator neutrinos with harder spectra required the recalculation of the charge-exchange strength function of the ${}^{127}$I nucleus and to calculate the capture cross sections for such hard neutrinos \cite{Lutostansky:2025pkh}.

Iodine is one of the heaviest targets for which processes of the inelastic neutrino--nucleus interaction have been measured \cite{PhysRevLett.131.221801}. In addition, the neutrino capture cross section on ${}^{127}$I at energies up to $E_\nu \approx$ 55~MeV and the cross sections for the emission of zero, one, or more neutrons were measured in \cite{PhysRevLett.131.221801}. 
The authors of \cite{PhysRevLett.131.221801, PhysRevD.106.032003, PhysRevC.109.035802} calculated cross sections for the capture of acceleration neutrinos on the ${}^{127}$I nucleus and it was noted in \cite{PhysRevC.109.035802} that such calculations of these cross sections on this nucleus are very important, since experimental measurements of these cross sections for ground-based neutrino/antineutrino sources in the low and intermediate energy ranges are very limited.

\begin{figure}[ht]
	\centering
	\includegraphics[width=\linewidth]{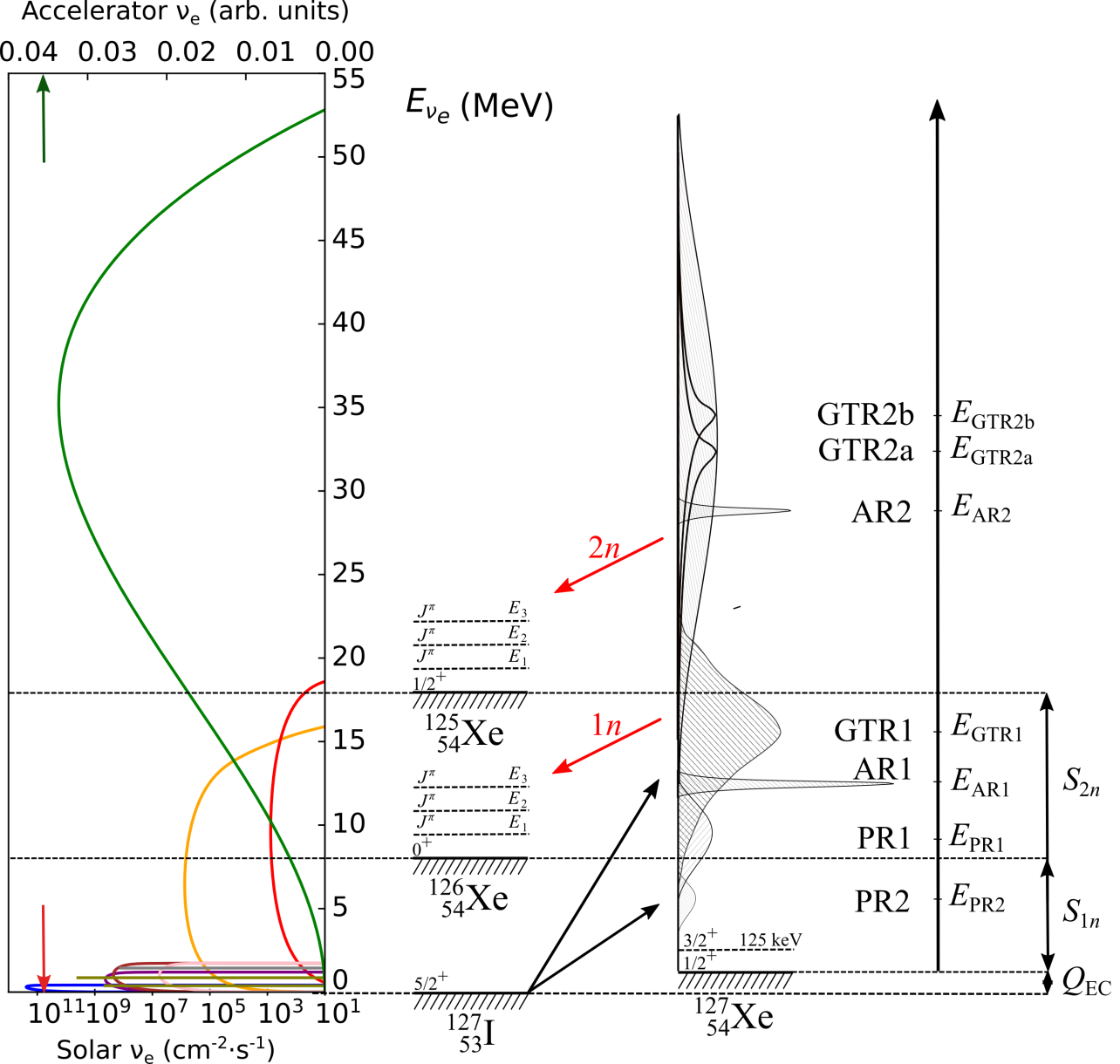}
	\caption{Diagram of charge-exchange excitations in the ${}^{127}$I($\nu_{e}$, $e^{-}$)${}^{127}$Xe reaction with decays of high-lying excitations in ${}^{127}$Xe into ${}^{126}$Xe and ${}^{125}$Xe isotopes with the emission of one and two neutrons with the corresponding energies $S_{1n}$ and $S_{2n}$, respectively. In the left part is the spectra of solar (up to 18~MeV) and the green line is accelerator neutrinos up to 55~MeV.}
	\label{fig1}
\end{figure}

The spectrum of accelerator neutrinos extends up to $\sim$55~MeV \cite{PhysRevD.106.032003, PhysRevC.109.035802}; i.e., it is much harder than the spectrum of solar neutrinos extending up to $\sim$20~MeV \cite{Bahcall_book} (see Fig.~\ref{fig1}), and this implies knowledge of the spectrum of charge-exchange excitations and the strength function $S(E)$ up to energies that have not yet been taken into account either in experiments or in calculations. The calculations presented in this work show that double AR-2 and GTR-2 resonances associated with single-particle spin--isospin shell transitions with changes in the main quantum number should exist above the well-known analog, AR-1, and Gamow--Teller, GTR-1, resonances (see Fig.~\ref{fig1}).

Figure~\ref{fig1} shows the scheme of charge-exchange excitations in the ${}^{127}$I($\nu_{e}$, $e^{-}$)${}^{127}$Xe reaction with the decay of high-lying excitations into the ${}^{126}$Xe and ${}^{125}$Xe isotopes with the emission of one and two secondary neutrons during the excitations of the ${}^{127}$Xe daughter nucleus with energies above $S_{1n} = 7.246$~MeV and $S_{2n} = 17.264$~MeV, respectively \cite{Wang_2021}. 
The energy threshold $Q_{\text{EC}} = 662.3$~keV \cite{Wang_2021} is determined from the mass difference of neighboring isobar nuclei. 
Its effect is small for accelerator neutrinos and is significant for solar neutrinos, since it includes the influence of low energy neutrinos \cite{LutostanskyEtAl:2020_JETP_lett, LUTOSTANSKY2022136905}, whose number is orders of magnitude larger. 
Two giant Gamow--Teller resonances, GTR-1 and GTR-2, as well as two analog resonances, AR-1 and AR-2, are marked. 
Lower-lying pygmy resonances PR are also indicated. All these resonances constitute the charge-exchange strength function $S(E)$ of the isotope ${}^{127}$I and make the main contribution to the cross sections for neutrino interaction with the ${}^{127}$I nucleus. 
The spectra of accelerator and solar neutrinos are also presented in the same energy scale, which makes it possible to determine which parts of the strength function $S(E)$ make the main contribution to the interaction of different types of neutrinos.

The only measurement of the strength function $S(E)$ was carried out long ago in 1999 in the ${}^{127}$I($p, n$)${}^{127}$Xe reaction \cite{Palarczyk_PhysRevC.59.500} up to an energy of 20~MeV. As it was noted in \cite{Palarczyk_PhysRevC.59.500}, the comparison with the experimental dependence of the strength function $S(E)$ showed that our earlier calculations of 1991 \cite{Lutostansky_Shulgina_PhysRevLett.67.430} had better predictive accuracy compared to other calculations \cite{Engel_Pittel_Vogel_PhysRevLett.67.426, Engel_Pittel_Vogel_PhysRevC.50.1702}.

\section{Resonant charge-exchange excitations of the ${}^{127}$I isotope}
The analysis of experimental data \cite{Palarczyk_PhysRevC.59.500} in comparison with our microscopic calculations \cite{Lutostansky_Shulgina_PhysRevLett.67.430, LutostanskyEtAl:2020_JETP_lett, LUTOSTANSKY2022136905} showed that resonance excitations make the main contribution to the interaction of solar neutrinos with the nuclei of the iodine detector.
The analysis of the interaction of this detector with harder accelerator neutrinos is given in this article.

As a result of processing experimental data on the ${}^{127}\mathrm{I}(p, n){}^{127}\mathrm{Xe}$ reaction \cite{Palarczyk_PhysRevC.59.500} up to an energy of 20~MeV \cite{LutostanskyEtAl:2020_JETP_lett, LUTOSTANSKY2022136905}, Gamow--Teller (GTR), analog (AR), and lower-lying (PR) pygmy resonances were identified. 
Performed microscopic calculations explained the nature and structure of these resonances. 
However, the previously obtained data on the strength function $S(E)$ turned out to be insufficient for the analysis of experiments with accelerator neutrinos with energies up to $\sim$55~MeV, and calculations were carried out by us up to the energy 60~MeV.

These new calculations within a microscopic approach using the Saxon--Woods model for the calculations of the one-particle structure showed that double GTR-2 and AR-2 resonances should be located above the Gamow--Teller and analog resonances, associated with the corresponding allowed one-particle transitions with selection rules $\Delta j = 0, \pm 1$, but with a change $\Delta n = 1$ in the main quantum number. These transitions in the ${}^{127}$I isotope are as follows: 
1) single-particle $(n-p)$ transitions with $\Delta j = 0$ that form the AR-2 resonance: $(2-3)s_{1/2}$, $(1-2)d_{3/2}$, $(1-2)d_{5/2}$, $(1-2)f_{7/2}$, $(1-2)f_{5/2}$, $(2-3)p_{3/2}$, and $(2-3)p_{1/2}$; 
2) single-particle $(n-p)$ transitions that form the GTR-2 resonance: transitions of two types -- with $\Delta j = +1$: $(2p_{1/2}-3p_{3/2})$, $(1d_{3/2}-2d_{5/2})$, $(1f_{5/2}-2f_{7/2})$, and transitions with $\Delta j = -1$: $(2p_{3/2}-3p_{1/2})$, $(1d_{5/2}-2d_{3/2})$, $(1f_{7/2}-2f_{5/2})$. Thus, the GTR-2 resonance must decay into two components with the spin--orbit splitting energy. However, these transitions should be suppressed by a phenomenological factor of $(E_{\mathrm{GTR2}} / E_{\mathrm{GTR1}})^{(2\Delta n + 1)}$ times.

Charge-exchange excitations of the ${}^{127}$I isotope were calculated in the microscopic theory of finite Fermi systems (TFFS) \cite{Migdal_book} as in \cite{LutostanskyEtAl:2020_JETP_lett, LUTOSTANSKY2022136905, Lutostansky:2025pkh}, but taking into account high-lying excitations up to the energy 60~MeV. The system of secular equations of the TFFS for the effective field was solved for allowed transitions with a local nucleon--nucleon interaction $F_{\omega}$ in the Landau--Migdal form \cite{Migdal_book}:
\begin{equation}
	F_\omega = C_0 (f_0^\prime + g_0^\prime(\boldsymbol{\sigma}_1 \boldsymbol{\sigma}_2)) (\boldsymbol{\tau}_1 \boldsymbol{\tau}_2) \delta(\boldsymbol{r}_1 - \boldsymbol{r}_2),
\end{equation}
where $C_0 = \left({d\rho}/{d\varepsilon_F}\right)^{-1} = 300~\text{MeV} \cdot \text{fm}^3$ ($\rho$ is the average density of nuclear matter) and $f_0^\prime$ and $g_0^\prime$ are the parameters, respectively, of the isospin-isospin and spin-isospin interactions of quasiparticles. In this work, we used the values $f_0^\prime = 1.351$ and $g_0^\prime = 1.214$, obtained in \cite{Lutostansky2020} from the analysis of calculated and experimental data on the energies of analog (38 nuclei) and Gamow--Teller (20 nuclei) resonances. Calculations within the TFFS with the Fayans functional should also be mentioned~\cite{Borzov2020}.

\begin{figure}[ht!]
	\centering
	\includegraphics[width=\linewidth]{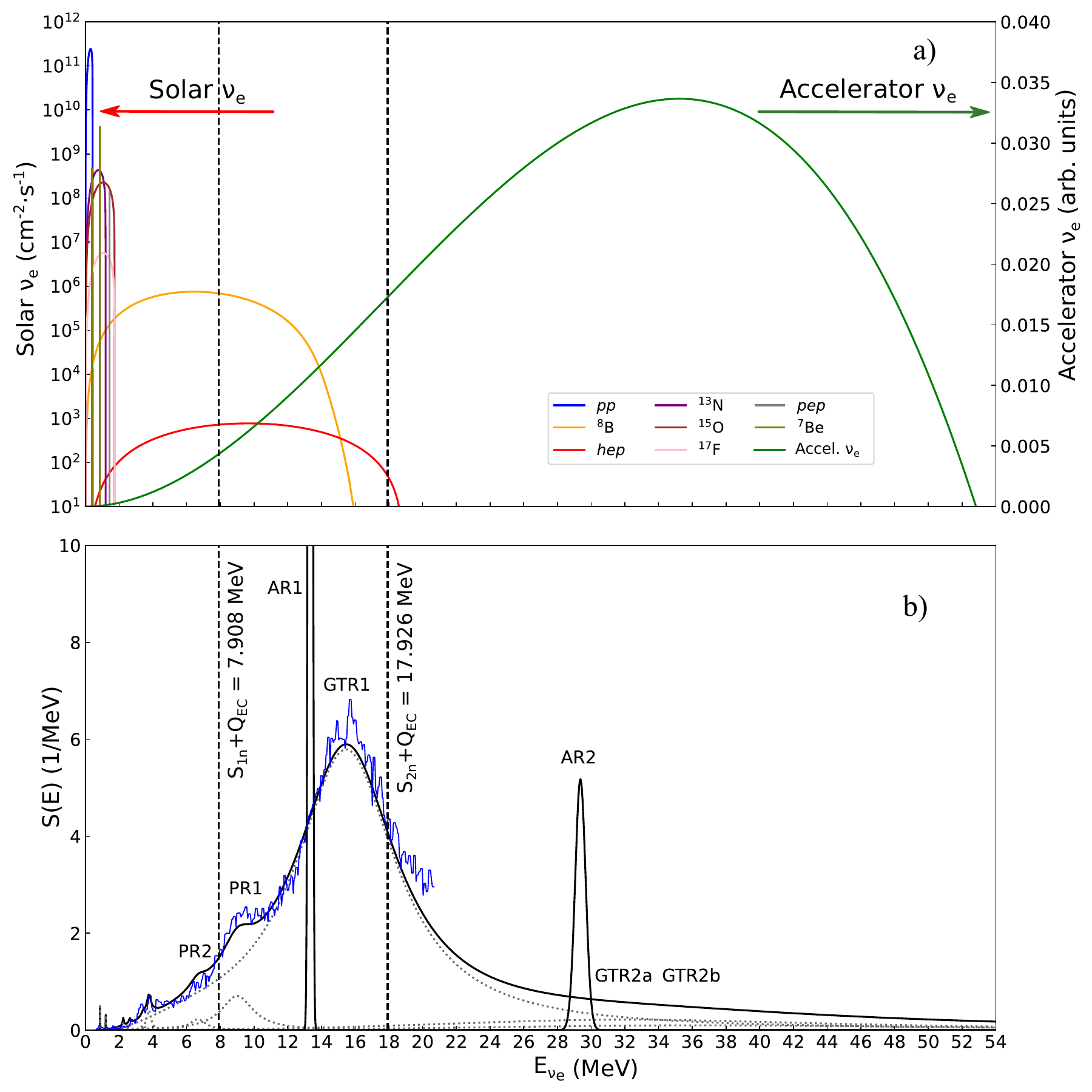}
	\caption{(a) Energy spectra of solar \cite{Bahcall_book} and accelerator \cite{PhysRevD.106.032003} neutrinos. (b) Charge-exchange strength function $S(E)$ for the excitation of Gamow--Teller and analog resonances in ${}^{127}$I. Blue solid lines present experimental data on the ${}^{127}$I($p, n$)${}^{127}$Xe reaction \cite{Palarczyk_PhysRevC.59.500}, black solid line presents our calculation within the theory of finite Fermi systems (TFFS), and dotted lines correspond to the GTR-1, GTR-2, PR1, PR2, and PR3 resonances.}
	\label{fig2}
\end{figure}

The calculations gave the matrix elements and the following energies of charge-exchange excitations of the ${}^{127}$I isotope up to 60~MeV: $E_{\text{GTR1}} = 14.8$~MeV, $E_{\text{PR1}} = 8.34$~MeV, $E_{\text{PR2}} = 6.0$~MeV, $E_{\text{AR1}} = 12.68$~MeV, $E_{\text{GTR2a}} = 30.8$~MeV, $E_{\text{GTR2b}} = 34.3$~MeV, and $E_{\text{AR2}} = 28.7$~MeV. 

Gamow--Teller (GTR), analog (AR) and pygmy (PR) resonances are clearly identified, for which the strength function $S(E)$ envelope curves with broadening according to the Breit--Wigner formulas were constructed, as in \cite{LutostanskyEtAl:2020_JETP_lett, LUTOSTANSKY2022136905} (see Fig.~\ref{fig2}b). Accordingly, the partial contributions were broadened as:
\begin{equation}
	S_i(E) = M_i^2 \frac{\Gamma_i^2}{(E-E_i)^2-\Gamma_i^2} \phi(E),
\end{equation}
where $\phi(E) = 1 - \exp[-(E/\Gamma_i)^2]$ is the form factor \cite{Migdal_book}, determining the value $S(E) = 0$ (or $\phi(E) = 0$) at the energies $E \le Q$, that is less than the threshold energy of the reaction (\ref{eq:1}).

According to \cite{Migdal_book}, the width was chosen in the form:
\begin{equation}
	\Gamma(E) = \alpha E^2 + \beta E^3 + \ldots
\end{equation}
In calculating $S(E)$, it is sufficient to use only the first term of the series with $\alpha \approx \epsilon_F^{-1}$, which effectively takes into account three-quasiparticle configurations. We used the value $\alpha = 0.018$~MeV$^{-1}$, obtained from the averaged experimental widths of GTR resonances. The low-lying discrete states were investigated separately.

Figure~\ref{fig2} shows: (a) the spectra of solar \cite{Bahcall_book} and accelerator \cite{PhysRevD.106.032003} neutrinos and (b) the experimental \cite{Palarczyk_PhysRevC.59.500} and calculated charge-exchange strength functions $S(E)$ of the ${}^{127}$I isotope (observed in ${}^{127}$Xe). Such a combination of two graphs allows to determine the contributions from various segments of the resonant structure of the strength function $S(E)$ to the interaction of neutrinos with the detector. It can be seen that it is sufficient to know the strength function $S(E)$ up to an energy of 20~MeV to calculate and analyze the interaction of solar neutrinos with ${}^{127}$I nuclei. However, to calculate and analyze the interaction of accelerator neutrinos up to an energy of 55~MeV, the strength function must be known up to such high energies.

Thus, the presented calculations of the strength function $S(E)$ of the ${}^{127}$I isotope at high energies make it possible to calculate the interaction of accelerator neutrinos with the iodine detector. It is also possible to analyze the contribution of different resonances to the energy ranges limited by the neutron separation energies in the ${}^{127}$Xe daughter nucleus, and to compare it with experimental data and other calculations (see below).

\section{Normalization of the strength function and \textit{quen\-ching} effect}

When describing the charge-exchange experimental and calculated strength functions $S(E)$ of the isotope ${}^{127}$I (observed in ${}^{127}$Xe) -- presented in Fig.~\ref{fig2}b, an important issue is the normalization of the $S(E)$-function. Thus, the experimental data for ${}^{127}$I were obtained in the reaction ${}^{127}$I($p, n$)${}^{127}$Xe \cite{Palarczyk_PhysRevC.59.500} and the charge-exchange strength function $S(E)$ was obtained up to the excitation energy $E_{max} = 20$~MeV. It was found that the total sum of the GT matrix elements $B(\textrm{GT})$ up to the energy of 20~MeV is equal to 53 with statistical uncertainty $\pm 0.22$ and systematic error $(+3.32, -19.47)$. That is $\approx$85\% of the maximum value of $3(N - Z) = 63$, which is given by the sum rule for GT excitations of the ${}^{127}$I nucleus \cite{Palarczyk_PhysRevC.59.500}. This means that there is a deficit in the sum rule for GT excitations -- the suppression effect or \textit{quenching} effect.

The observed deficit in the sum rule for GT excitations is associated with the \textit{quenching} effect \cite{ARIMA1999260} or with a violation of the normalization of GT matrix elements. Thus, according to the well-known sum rule, for GT-transitions, the normalization has the form:
\begin{equation}
	\label{eq:Mi_summ}
	\sum M_i^2 = \sum B_i(\textrm{GT}) = q [3(N-Z)] = e_q^2 [3(N-Z)] = \int_{0}^{E_{max}} S(E) dE = I(E_{max}),
\end{equation}
where $E_{max}$ is the maximum energy taken into account in the calculation or in the experiment and $S(E)$ is the charge-exchange strength function. The parameter $q < 1$, in Eq.~(\ref{eq:Mi_summ}) determines the \textit{quenching} effect (deficit in the sum rule) and at $q = 1$, which corresponds to the maximum value $\sum M_i^2 = \sum B_i(\textrm{GT}) = 3(N-Z)$ in Eq.~(\ref{eq:Mi_summ}). Within the TFFS framework, $q = e_q^2$, where $e_q$ is an effective charge \cite{Migdal_book}. As was shown by A.~B.~Migdal \cite{Migdal_1957}, the effective charge should not exceed unity; for Fermi transitions, we have $e_q(\mathrm{F}) = 1$, while, for GT transitions, $e_q(\mathrm{GT}) = 1-2\zeta_S$ (see \cite{Migdal_book}, p.~223), where $\zeta_S$, $0 < \zeta_S < 1$, is an empirical parameter. Thus, we see that, in the case of I $\to$ Xe transitions considered here, the effective charge $e_q = e_q(\mathrm{GT})$ is a parameter extracted from the experimental data.

\begin{figure}[ht!]
	\centering
	\includegraphics[width=\linewidth]{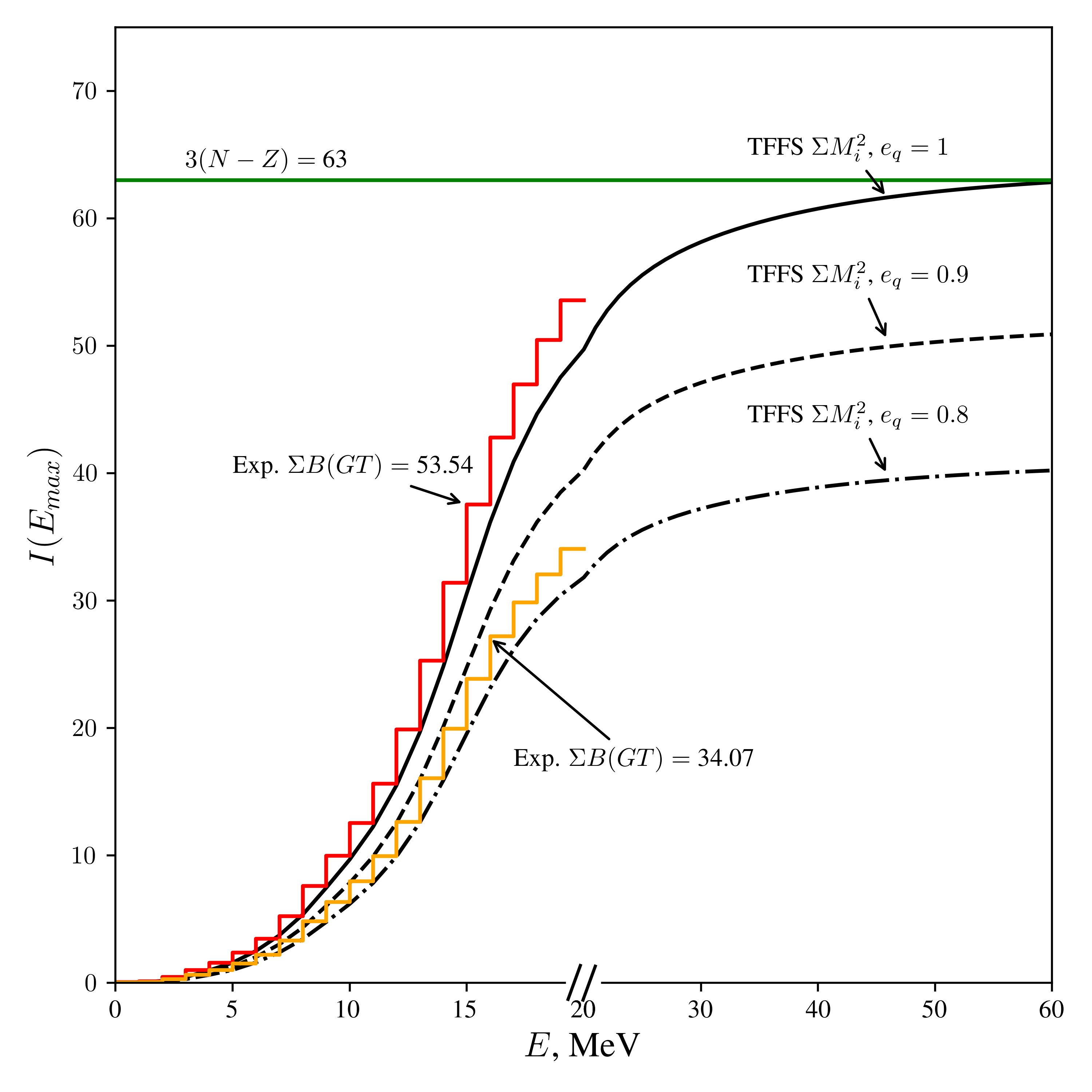}
	\caption{Integral $I(E_{max})$ given by Eq.~(\ref{eq:Mi_summ}) versus the energy $E_{max}$ for the ${}^{127}$Xe isotope: (steps) experimental data from \cite{Palarczyk_PhysRevC.59.500}, (solid line) calculation with $e_q = 1.0$, dotted line -- calculation with $e_q = 0.9$, (dash-dotted line) calculation with $e_q = 0.8$, and (horizontal straight line) sum-rule value of $3(N - Z) = 63$. The scale on the abscissa axis changes after 20~MeV.}
	\label{fig3}
\end{figure}

The experimental value of the \textit{quenching} parameter may change greatly from one nucleus to another \cite{Palarczyk_PhysRevC.59.500, PhysRevLett.54.2325, Lutostansky_Tikhonov2018, Lutostansky:2019iri} -- for example, from $q = 0.67$ or $67\pm8\%$ for ${}^{98}$Mo \cite{PhysRevLett.54.2325} to $q = 0.85$ or $85\%$ in the case of the ${}^{127}$I nucleus \cite{Palarczyk_PhysRevC.59.500}. In this case, processing the results of charge-exchange reactions can give significant errors in the final value of $\sum B_i(\textrm{GT})$, as can be seen in the example of ${}^{127}$I, for which $I(E_{max}) \approx 85\%$ of the maximum value $3(N - Z) = 63$ with systematic uncertainty of $(+5.27, -30.9)\%$. This error is due both to the low energy resolution in the $(p,n)$ reaction and to the uncertainty in the data analysis at high nuclear excitation energies, where transitions to quasi-free states (QFC) occur \cite{PhysRevC.51.526}. This uncertainty significantly complicates both the comparison of theoretical calculations of the squares of matrix elements with experimental data, and the calculation of cross sections and neutrino capture rates for nuclei. A new measurement of the $S(E)$ strength function with high energy resolution is needed, for example for the ${}^{127}$I(${}^{3}$He, $t$)${}^{127}$Xe reaction at higher energies.

Figure~\ref{fig3} shows the integral $I(E_{max})$ in Eq.~(\ref{eq:Mi_summ}) as a function of $E_{max}$ for the daughter isotope ${}^{127}$Xe. Colored lines -- $S(E)$, obtained from the $(p,n)$ reaction \cite{Palarczyk_PhysRevC.59.500}, black lines -- $S(E)$, calculated in the TFFS with different values of the parameter $e_q$. The red line corresponds to the upper limit of the experimental sum $\sum B_i(\textrm{GT})=53.54$, the yellow line -- the lower limit $\sum B_i(\textrm{GT})=34.07$, which were obtained up to the energy $E_{max} = 20$~MeV. The average between the two experimental values is best described by the theoretical $S(E)$, with the effective charge value $e_q = 0.9$ ($q = 0.81$) -- the black dashed-dotted line in Fig.~\ref{fig3}. It can be seen that, in this case, the best description of the experimental data is obtained in calculations up to energy $E_{max}$ with the effective-charge value of $e_q = 0.9$ ($q = 0.81$). For other nuclei, however, the calculated values of $e_q$ differ from $0.9$, predominantly taking smaller values \cite{PhysRevLett.54.2325, Lutostansky_Tikhonov2018}. Mostly, this is characteristic of nuclei lighter than ${}^{127}$I and is partly due to the disregard of high-lying (above GTR) excitations in the experiment that are formed by single-particle transitions in which $\Delta n = 1, 2$.

At excitation energies above 20~MeV, the main contribution to the strength function $S(E)$ comes from states formed by transitions with $\Delta n = 1, 2$ and the right-hand tail of the GTR. The contribution of states with $\Delta n = 1, 2$ is $\Delta I = 3.8\%$ of the total integral $I(E_{max})$.

Thus, the calculation results with $e_q = 0.9$ ($q = 0.81$) have best fit to experimental data with the energies up to 20~MeV and suggest the presence of a \textit{quenching} suppression effect at higher energies. It is difficult to say how much this corresponds to experiment, since at energies exceeding the detachment energy of one or more neutrons, other effects that determine the suppression of the neutrino capture process may be possible.

\section{Cross sections and capture rates of accelerator neutrinos by the iodine detector}

The dependence of the ${}^{127}$I($\nu_{e}$, $e^{-}$)${}^{127}$Xe reaction cross section $\sigma(E)$ on the energy of the incident neutrino $E_{\nu}$ is given by the expression:
\begin{equation}
	\begin{split}
		\sigma\left( E_{\nu} \right) &= \frac{\left( G_{F}g_{A} \right)^{2}}{\pi c^{3}\hbar^{4}} 
		\int_{0}^{E_{\nu} - Q} E_{e} p_{e} F\left( Z,A,E_{e} \right) S(x)\,dx \\
		E_{e} &= E_{\nu} - Q - x + m_{e}c^{2}, \quad 
		cp_{e} = \sqrt{E_{e}^{2} - \left(m_{e}c^{2}\right)^{2}}
	\end{split}
\end{equation}
where $F(Z, A, E_e)$ is the Fermi function, $S(E)$ is the strength function, $G_F/(\hbar c)^3 = 1.1663787(6) \times 10^{-5}$~GeV$^{-2}$ is the weak interaction constant and $g_A = -1.2754 \pm 0.0013$ is the axial-vector constant from \cite{PhysRevD.110.030001}.

The Fermi function describes the interaction of an emitted electron with the Coulomb field of the nucleus, and its value directly affects the cross section $\sigma(E)$. An accurate calculation of $F(Z, A, E_e)$ is quite laborious and requires the inclusion of corrections for the size of the nucleus and the charge distribution in it, screening of the charge by orbital electrons, and other effects. The existing tables of this function (see \cite{Dzhelepov_1972, Janecki_1969}), which were obtained by the numerical solution of the Dirac equation in the Coulomb field of the nucleus, cover the emitted electron energies only up to $E_e \approx 10$--$15$~MeV, which are significantly less than the energy range of electrons emitted due to the capture of accelerator neutrinos. It is also noteworthy that corrections to the shape of the beta spectrum and the Fermi function were considered analytically in \cite{fermi_function_2017}. We showed recently in \cite{Fazliakhmetov:2023ast} that different calculations of the Fermi function for $E_e < 10$--$15$~MeV give values that differ up to $\approx$15\%.

In this work, a linear extrapolation of the Fermi function values from \cite{Dzhelepov_1972} was used. The values of the function $p_e F(Z, A, E_e)$ were extrapolated, since this function increases almost linearly in the range from 8~MeV to the boundary energy from the tables presented in \cite{Dzhelepov_1972}. The current approximations of the behavior of the Fermi function at high electron energies \cite{PhysRevC.57.2004} are inconsistent with the calculated $F(Z, A, E_e)$ values (e.g., \cite{Dzhelepov_1972, Janecki_1969}) for $E_e < 10$~MeV \cite{PhysRevC.103.044604}. Thus, tabular values of the Fermi function for the energy range of electrons emitted due to the capture of accelerator neutrinos are currently absent.

The neutrino capture cross sections averaged over the energy spectrum were calculated by the formula:
\begin{equation}
	\langle \sigma \rangle = \int_{0}^{E_{\nu}^{\max}} \sigma(E_{\nu}) p(E_{\nu}) dE_{\nu}
\end{equation}
where $p(E_{\nu})$ is the spectrum of electron neutrinos from $\pi^{+}$ meson decays, the analytical form of which is taken from \cite{PhysRevC.109.035802}.

The spectrum-averaged neutrino capture cross sections calculated in this work taking into account the strength function $S(E)$ obtained in the microscopic TFFS \cite{Migdal_book} -- $\langle \sigma \rangle_{\text{TFFS}}$, are presented in Table~\ref{table1} in comparison with the calculated values and experimental measurements of other scientific groups. 
\begin{table}[h]
	\centering
	\caption{Spectrum averaged cross sections for neutrino capture in the units of $10^{-40}$~cm$^2$: $\langle \sigma(0n) \rangle$ -- is the cross section for transitions only to bound states of ${}^{127}$Xe without neutron emission, $\langle \sigma(\ge 1n) \rangle$ -- is the cross section for transitions to ${}^{126}$Xe and ${}^{125}$Xe with the emission of one ($\langle \sigma(1n) \rangle$) or two ($\langle \sigma(2n) \rangle$) neutrons, $\langle \sigma \rangle_{\text{COHERENT}}$ -- is the experimental cross section obtained by the COHERENT collaboration \cite{PhysRevLett.131.221801}. $\langle \sigma \rangle_{\text{MARLEY}}$ -- is the cross section obtained in the COHERENT calculations using the MARLEY Monte Carlo simulation (from the appendix in \cite{PhysRevLett.131.221801}), $\langle \sigma \rangle_{\text{LAMPF}}$ -- is the cross section measured on the prototype of the iodine radiochemical solar neutrino detector in Los Alamos \cite{PhysRevC.68.054613}, and $\langle \sigma \rangle_{\text{TFFS}}$ -- is the cross section obtained in our calculations using $S(E)$, obtained in the microscopic theory of finite Fermi systems \cite{Migdal_book}.}
	\begin{adjustbox}{width=\linewidth}
		\begin{tabular}{|l|c|c|}
			\hline
			& $\langle \sigma(0n) \rangle$ & $\langle \sigma(\geq 1n) \rangle$ \\
			\hline
			$\langle \sigma \rangle_{\text{COHERENT}}$ & $5.2_{-3.1}^{+3.4}$ & $2.2_{-2.2}^{+3.5}$ \\
			\hline
			$\langle \sigma \rangle_{\text{MARLEY}}$ & $2.3_{-1.7}^{+0.2}$ & $\langle \sigma(1n) \rangle = 18.9_{-5.3}^{+1.0}$ \\
			& & $\langle \sigma(2n) \rangle = 0.8_{-0.4}^{+0.1}$ \\
			\hline
			$\langle \sigma \rangle_{\text{LAMPF}}$ & $2.84 \pm 0.91$(stat.) $\pm 0.25$(syst.) & $-$ \\
			\hline
			$\langle \sigma \rangle_{\text{TFFS}}$ & 1.94 & $\langle \sigma(1n) \rangle = 16.07$ \\
			& & $\langle \sigma(2n) \rangle = 2.88$ \\
			\hline
		\end{tabular}
	\end{adjustbox}
	\label{table1}
\end{table}
In Table~\ref{table1} and everywhere in this work, $\langle \sigma(0n) \rangle$ -- is the neutrino capture cross section for transitions to ${}^{127}$Xe without neutron emission in the nuclear excitation energy range $E_x < S_{1n}$, $\langle \sigma(1n) \rangle$ -- is the neutrino capture cross section for transitions to ${}^{126}$Xe with the emission of one neutron in the nuclear excitation energy range $S_{1n} < E_x < S_{2n}$, and $\langle \sigma(2n) \rangle$ -- is the neutrino capture cross section for transitions to ${}^{125}$Xe with the emission of two neutrons in the nuclear excitation energy range $E_x > S_{2n}$. These ranges were chosen for convenient comparison with the results of other groups reported in \cite{PhysRevLett.131.221801, PhysRevC.68.054613} and \cite{PhysRevC.109.035802}.

Our calculated section values without neutron emission $\langle \sigma(0n) \rangle$ are in good agreement with both the Monte Carlo MARLEY simulation results (see Appendix in \cite{PhysRevLett.131.221801}) and the experimental measurements obtained in two different experiments reported in \cite{PhysRevLett.131.221801, PhysRevC.68.054613}. The situation for the cross sections $\langle \sigma(\ge 1n) \rangle$ for transitions to ${}^{126}$Xe and ${}^{125}$Xe with the emission of one and two neutrons, respectively, is different: the COHERENT collaboration measurements -- $\langle \sigma \rangle_{\text{COHERENT}}$ \cite{PhysRevLett.131.221801} differ significantly from both the MARLEY results -- $\langle \sigma \rangle_{\text{MARLEY}}$ (see Appendix to \cite{PhysRevLett.131.221801}) and -- $\langle \sigma \rangle_{\text{TFFS}}$ -- our calculated values. At the same time, our values are in good agreement with the MARLEY results.

A similar discrepancy between the experimentally measured and theoretically calculated cross sections $\langle \sigma(\ge 1n) \rangle$ for lead was found by the same authors from the COHERENT collaboration \cite{PhysRevD.108.072001}. The reasons for this discrepancy between experiment and theory are not fully understood. The authors of \cite{PhysRevLett.131.221801} proposed a decrease in the axial interaction constant $g_A$ in the nucleus (\textit{quenching}-effect) as a possible explanation. The theoretical calculations performed under this assumption in \cite{PhysRevC.109.035802} show good agreement with the COHERENT measurements for $\langle \sigma(\ge 1n) \rangle$, but the cross section $\langle \sigma(0n) \rangle = 6.1 \times 10^{-40}~\text{cm}^2$ obtained in \cite{PhysRevC.109.035802} differ significantly from $\langle \sigma(0n) \rangle$ measured in Los Alamos \cite{PhysRevC.68.054613}.
\begin{table}[h]
	\centering
	\caption{Contributions of the AR1, AR2, GTR1, GTR2, PR1, and PR2 resonances to the neutrino capture cross section $\langle \sigma \rangle_{\text{TFFS}}$ for different nuclear excitation energy ranges $E_x$. The cross-section values are given in units of $10^{-40}$~cm$^2$.}
	\begin{adjustbox}{width=\linewidth}
		\begin{tabular}{|l|c|c|c|}
			\hline
			& $E_x < S_{1n}$ & $S_{1n} < E_x < S_{2n}$ & $E_x > S_{2n}$ \\
			\hline
			${\langle \sigma \rangle}_{\text{TFFS}}$ & 1.94 & 16.07 & 2.88 \\
			\hline
			${\langle \sigma \rangle}_{\text{TFFS}}$ contribution of AR1 & 0 & 5.42 & 0 \\
			\hline
			${\langle \sigma \rangle}_{\text{TFFS}}$ contribution of AR2 & 0 & 0 & 0.23 \\
			\hline
			${\langle \sigma \rangle}_{\text{TFFS}}$ contribution of GTR1 & 1.36 & 9.81 & 2.28 \\
			\hline
			${\langle \sigma \rangle}_{\text{TFFS}}$ contribution of GTR2 & 0.02 & 0.16 & 0.36 \\
			\hline
			${\langle \sigma \rangle}_{\text{TFFS}}$ contribution of PR1 & 0.24 & 0.65 & 0 \\
			\hline
			${\langle \sigma \rangle}_{\text{TFFS}}$ contribution of PR2 & 0.16 & 0.02 & 0 \\
			\hline
		\end{tabular}
		\label{table2}
	\end{adjustbox}
\end{table}

A significant problem for theoretical calculations of the neutrino capture cross section is the lack of experimental data on charge-exchange GT states for nuclear excitation energies above 20~MeV. As noted above, the only available data are measurements for the ${}^{127}$I($p, n$)${}^{127}$Xe reaction in 1999 \cite{Palarczyk_PhysRevC.59.500}. For the current studies concerning accelerator neutrinos with energies up to 55~MeV and higher, the results of the experiment reported in \cite{Palarczyk_PhysRevC.59.500} are insufficient and a more accurate experiment is needed, e.g., on the ${}^{127}$I(${}^{3}$He, $t$)${}^{127}$Xe reaction at higher energies.

Table~\ref{table2} presents the contributions of the AR1, AR2, GTR1, GTR2, PR1, and PR2 resonances to the neutrino capture cross section in three nuclear excitation energy ranges: (i) $E_x < S_{1n}$ without neutron emission, (ii) $S_{1n} < E_x < S_{2n}$ with the emission of one neutron, and (iii) $E_x > S_{2n}$ with the emission of two neutrons. In all three ranges, the dominant contribution comes from the GTR1 resonance: 70\% at $E_x < S_{1n}$, 61\% at $S_{1n} < E_x < S_{2n}$ and 79\% at $E_x > S_{2n}$. The PR1 and PR2 resonances make a significant contribution of 20\% to the cross section only in the range $E_x < S_{1n}$, and their contribution at $S_{1n} < E_x < S_{2n}$ is only about 4\%. The AR1 resonance makes a contribution of 34\% in the nuclear excitation energy range $S_{1n} < E_x < S_{2n}$, the contribution of the AR2 resonance to the cross section is about 8\% at $E_x > S_{2n}$, where the contribution from the GTR2 resonance to the cross section is 12.5\%.

Thus, the calculations show (see Table~\ref{table2}) that the determining contribution in all three energy ranges comes from the GTR1 resonance, and the AR1 resonance makes a significant contribution at the energies $S_{1n} < E_x < S_{2n}$. The contributions of the secondary GTR2 and AR2 resonances are noticeable only at high energies $E_x > S_{2n}$ and are not determining.

\section{Conclusion}

The interaction of high-energy neutrinos from the SNS accelerator at the Oak Ridge National Laboratory (United States) has been studied with a promising new ${}^{127}$I detector. The resonance structure of the charge-exchange strength function $S(E)$ has been calculated within the microscopic theory of finite Fermi systems taking into account high-lying GTR1, AR1, PR1, and PR2 resonances and secondary GTR-2 and AR-2 resonances. Using the obtained strength function $S(E)$, the spectrum-averaged cross sections $\langle \sigma(E_{\nu}) \rangle$ for the capture of electron neutrinos with energies up to 55~MeV from the SNS accelerator in three nuclear excitation energy ranges have been calculated.

Our results for cross sections without neutron emission $\langle \sigma(0n) \rangle$ are in good agreement with experimental measurements in \cite{PhysRevLett.131.221801, PhysRevC.68.054613}. The calculated cross sections for events with one or more neutrons $\langle \sigma(\ge 1n) \rangle$ are consistent with the MARLEY Monte-Carlo simulation results, but differ from the measurements of the COHERENT Collaboration \cite{PhysRevLett.131.221801}. This discrepancy can be explained both by new effects in nuclear theory and, partially, by the uncertainty in the calculation associated with the extrapolation of the Fermi function values from \cite{Janecki_1969}, and the lack of experimental data on charge-exchange GT states for the excitation energies of the nucleus ${}^{127}$Xe above 20~MeV. A new measurement of the strength function $S(E)$ with a high energy resolution is needed, e.g., on the ${}^{127}$I(${}^{3}$He, $t$)${}^{127}$Xe reaction at higher energies.

The contributions of resonances to the neutrino capture cross section have been estimated in three nuclear excitation energy ranges: up to the neutron separation energy $E_x < S_{1n}$, above the separation energy of one neutron and below the separation energy of the second neutron $S_{1n} < E_x < S_{2n}$, and above the separation energy of two neutrons $E_x > S_{2n}$. The significant effect of the Gamow--Teller resonance GTR1 and the analog resonance AR1 on the energy dependence of the cross section $\sigma(E)$ at the energies $E_x < S_{2n}$ has been demonstrated. The new GTR2 and AR2 resonances insignificantly affect the energy dependence of the cross section $\sigma(E)$ at the energies $E_x > S_{2n}$.

Thus, the comparison of the cross sections for the interaction of accelerator neutrinos calculated by different methods with experimental data has shown coincidence at energies below the neutron separation threshold $E_x < S_{1n}$ and strong discrepancy at higher energies, which is difficult to explain. A new measurement of cross sections at energies $E_x > S_{1n}$ is needed.

\section{Acknowledgments}

We are grateful to W. C. Haxton, M.D. Skorokhvatov, A.M. Konovalov, I. N. Borzov and S.V. Tolokonnikov for stimulating discussions and assistance in this work.

\printbibliography

\end{document}